\documentclass[aps,twocolumn,longbibliography] {revtex4-1}

\usepackage{bbm}

\usepackage[english]{babel}

\makeatletter
\def\bbl@set@language#1{%
	\edef\languagename{%
		\ifnum\escapechar=\expandafter`\string#1\@empty
		\else\string#1\@empty\fi}%
	\@ifundefined{babel@language@alias@\languagename}{}{%
		\edef\languagename{\@nameuse{babel@language@alias@\languagename}}%
	}%
	\select@language{\languagename}%
	\expandafter\ifx\csname date\languagename\endcsname\relax\else
	\if@filesw
	\protected@write\@auxout{}{\string\select@language{\languagename}}%
	\bbl@for\bbl@tempa\BabelContentsFiles{%
		\addtocontents{\bbl@tempa}{\xstring\select@language{\languagename}}}%
	\bbl@usehooks{write}{}%
	\fi
	\fi}
\newcommand{\DeclareLanguageAlias}[2]{%
	\global\@namedef{babel@language@alias@#1}{#2}%
}
\makeatother

\DeclareLanguageAlias{en}{english}

\usepackage{amsmath,graphicx,amsfonts}

\usepackage[colorlinks=true,linkcolor=blue,urlcolor=blue,citecolor=blue,breaklinks=true]{hyperref}

\usepackage[utf8]{inputenc}
\DeclareUnicodeCharacter{03BC}{\mbox{$\mu$}}
\DeclareUnicodeCharacter{2009}{ }

\def\wt{\widetilde}

\def\la{\langle}
\def\ra{\rangle}

\def\da{\dagger}

\def\da{\dagger}

\date\today
\begin{document}
	
\title{\mbox{Fluctuation-induced potential for an impurity in a semi-infinite one-dimensional Bose gas}}
\author{Benjamin Reichert}
\author{Aleksandra Petkovi\'{c}}
\author{Zoran Ristivojevic}
\affiliation{Laboratoire de Physique Th\'{e}orique, Universit\'{e} de Toulouse, CNRS, UPS, 31062 Toulouse, France}
	
\begin{abstract}
We consider an impurity in a semi-infinite one-dimensional system of weakly-interacting bosons. We calculate the interaction potential for the impurity due to the end of the system, i.e., the wall. For local repulsive (attractive) interaction between the impurity and the Bose gas, the interaction potential is attractive (repulsive). At short distances from the wall it decays exponentially crossing over into a universal $1/r^2$ behavior at separations $r$ above the healing length. Our results can also be interpreted as a Casimir-like interaction between two impurities, where one of them is infinitely strongly coupled to the Bose gas. We discuss various scenarios for the induced interaction between the impurities using the scattering approach. We finally address the phenomenon of localization of the impurity near the wall. In the paper we mainly study the case of a static impurity, however the universal $1/r^2$ interaction also holds for a slowly moving impurity.
\end{abstract}

\maketitle

\section{Introduction}

In classical electrodynamics, a charged particle experiences a force when placed in an electric field of another charge as described by Coulomb's law. Similarly, a charged particle is attracted by a metallic wall, which can be explained as the interaction with the image charge of the opposite sign \cite{Landau_vol8}.  For a ground-state atom in front of a conducting wall, Coulomb's law is not directly applicable since it is neutral. However, the atom possesses a fluctuating dipole and  it is attracted toward the wall due to the interaction with the induced image charge distribution. At short distances $r$ from the wall, the interaction potential behaves as $1/r^3$ \cite{lennard-jones_processes_1932}. Due to the retardation effects the potential crosses over into a $1/r^4$ law at long distances, as first shown by Casimir and Polder \cite{casimir_influence_1948}. Their result has been experimentally confirmed during recent years \cite{sukenik_measurement_1993,landragin_measurement_1996,shimizu_specular_2001,druzhinina_experimental_2003,pasquini_quantum_2004,harber_measurement_2005,bender_direct_2010}.

The latter example belongs to a wider class of fluctuation-induced phenomena where external bodies modify the fluctuations of the surrounding medium  \cite{kardar_friction_1999,bordag_advances_2009}. As a result, such perturbing objects experience an induced interaction, which could lead to qualitatively new effects. A well known example is the formation of Cooper pairs of electrons in the lattice of ions \cite{cooper_bound_1956}. Another one is the formation of bipolarons, which represent bound states of two quasiparticle polarons, and occurs, e.g., in ionic crystals \cite{vinetskii_bipolar_1961} or Bose-Einstein condensates \cite{camacho-guardian_bipolarons_2018}. 

The effect of quantum fluctuations, while being important for the above-mentioned cases, becomes particularly enhanced in low-dimensional systems. One therefore expects the most pronounced effects of the induced interaction to take place in such environments. In the following we consider a one-dimensional quantum liquid as a host medium, where the density fluctuations provide the leading mechanism that induces the interaction between external bodies (impurities). Additionally, various experimental realizations of these systems \cite{palzer_quantum_2009,bakr_quantum_2009,weitenberg_single-spin_2011,catani_quantum_2012,fukuhara_quantum_2013} compel us to understand the problem on theoretical grounds.

The induced interaction between impurities in one-dimensional quantum liquids is studied in several works \cite{recati_casimir_2005,fuchs_oscillating_2007,wachter_indirect_2007,kolomeisky_casimir_2008,yu_casimir_2009,dehkharghani_interaction-driven_2017,pavlov_phonon-mediated_2018,schecter_phonon-mediated_2014,reichert_casimir-like_2019,reichert_field-theoretical_2019}. In repulsively interacting fermion media, the smooth component of the induced, Casimir-like, interaction between heavy impurities scales as $1/r$ at large separations $r$ \cite{recati_casimir_2005,fuchs_oscillating_2007,wachter_indirect_2007}. In contrast to that, the long-range  interaction was not found for attractively interacting fermionic media, or equivalently, for repulsively interacting bosons in Refs.~\cite{recati_casimir_2005,fuchs_oscillating_2007,wachter_indirect_2007,dehkharghani_interaction-driven_2017}. However, a recent study of \citet{schecter_phonon-mediated_2014} reports on a long-range induced interaction in both, bosonic and fermionic media, which scales as $1/r^3$. In Refs.~\cite{reichert_casimir-like_2019,reichert_field-theoretical_2019} was calculated the induced interaction between heavy impurities in a weakly-repulsive Bose gas at arbitrary distances, which at large $r$ is in full agreement with Ref.~\cite{schecter_phonon-mediated_2014}. We finally notice that the authors of Ref.~\cite{schecter_phonon-mediated_2014} advocate a broad universality of their result, where the only exception is the case of infinite mass impurities in repulsive fermionic environments.

In this work we study the induced interaction between an impurity and a semi-infinite one-dimensional system of weakly interacting bosons. This realistic setup resembles the Casimir-Polder one consisting of a single atom and a wall.  Our problem can  also be interpreted as an interimpurity interaction, where one impurity is infinitely strongly coupled to the Bose gas and thus plays the role of the system end (wall). The theories \cite{schecter_phonon-mediated_2014,reichert_casimir-like_2019,reichert_field-theoretical_2019} do not apply to the latter case. We find that the induced interaction at large $r$ scales as $1/r^2$, where $r$ denotes the distance of the impurity from the wall. The latter interaction shows qualitative differences and decays slower than $1/r^3$ of Ref.~\cite{schecter_phonon-mediated_2014}, which immediately guarantees that the present setup will provide a stronger induced interaction effect. Our result is a genuine quantum effect, since the depletion of the classical, mean-field density of the Bose gas near its end (or near the strongly coupled impurity) quickly disappears beyond the healing length and thus it cannot lead to the long-range interaction. Interestingly, the impurity that repels the bosons can become localized near the wall. This occurs because  the two deeps in the density of bosons, one caused by the wall and the other by the impurity, tend to overlap in order to minimize the energy. The potential energy of this attraction  can overwhelm the kinetic energy of the localized impurity, leading to a bound state, as we discuss below.

\section{Model}

We study a one-dimensional  Hamiltonian
\begin{align}
H={}&\int_0^\infty dx\left(-\hat\Psi^\dagger\frac{\hbar^2\partial_x^2}{2m}\hat\Psi+\frac{g}{2}\hat\Psi^\dagger\hat\Psi^\dagger\hat\Psi\hat\Psi\right)\notag\\
&+G\hat\Psi^\dagger(r,t)\hat \Psi(r,t).
\label{eq:Hw}
\end{align}
The first line of Eq.~(\ref{eq:Hw}) describes a semi-infinite Bose gas with the contact repulsion of the strength $g$. By $m$ is denoted the mass of bosons. The remaining term accounts for the static impurity at the position $r$ that is locally coupled to the density of the system with the coupling constant $G$. The bosonic single particle field operators of Eq.~(\ref{eq:Hw}) satisfy the standard equal time commutation relation $[ \hat\Psi(x,t),\hat\Psi^\da(x',t)]=\delta(x-x')$. The Hamiltonian (\ref{eq:Hw}) should be supplemented by an additional boundary condition
\begin{align}
\hat\Psi(x=0,t)=0.
\label{eq:BC}
\end{align} 
Equation (\ref{eq:BC}) is compatible with the complete depletion of the boson density $\hat\Psi^\dagger \hat\Psi$ at $x=0$. This occurs either as a natural condition at the end of the semi-infinite system or due to an impurity at the origin that is infinitely strongly coupled to the Bose gas. Our goal in the following is to obtain analytically the dependence of the ground-state energy  of the Hamiltonian (\ref{eq:Hw}) on $r$ and hence find the induced interaction felt by the impurity.

\section{Equation of motion and its solution at weak interaction}

For the sake of simplicity of the presentation, we introduce the dimensionless quantities, where the length is measured in units of $\xi_\mu=\hbar/\sqrt{m\mu}$ while the time in units of $\hbar/\mu$. Here $\mu$ denotes the chemical potential. In these units, the field operator  becomes $\hat \Psi(x,t)=\sqrt{{\mu}/{g}}\,\hat \psi(X,T)e^{-i T}$, where $X$ and $T$ are dimensionless coordinate and time, respectively. The equation of motion for the field, $i\hbar \partial_t \hat \Psi=[\hat \Psi,H]$, becomes
\begin{align}
i\partial_T \hat\psi=\Big[ -\dfrac{\partial^2_X}{2} -1+\hat\psi^\da\hat\psi+\wt G\delta(X-R)\Big] \hat\psi.
\label{eq:eomw}
\end{align}
Here $R=r/\xi_\mu$ is the dimensionless distance of the impurity from the system end, while $G=\hbar \sqrt{\mu/m}\,\wt G$.

It is unknown how to solve Eq.~(\ref{eq:eomw}) exactly. However, we can study it using the perturbation theory at weak interaction between bosons. In this case we introduce a small dimensionless parameter $\gamma=mg/\hbar^2 n\ll 1$, where $n$ is the boson density. We then assume the field operator in the form   \cite{pitaevskii_bose-einstein_2003,sykes_drag_2009}
\begin{gather}
\hat\psi(X,T)=\psi_0(X)+\alpha\hat\psi_1(X,T)+\alpha^2\hat\psi_2(X,T)+\ldots \label{eq:wf_expw},
\end{gather}
where $\alpha=(\gamma g n/\mu)^{1/4}\ll 1$. Note that we are studying the problem in the grand canonical ensemble where $\mu$ is fixed, as will be discussed later. The function $\psi_0(X)$ represents the time independent wave function in the absence of fluctuations, whereas $\hat\psi_{1(2)}$ is the first (second) quantum correction. The expansion (\ref{eq:wf_expw}) is justified at $\ln (L/\xi_\mu)\ll 1/\sqrt{\gamma}$, where $L$ is the system size \cite{petrov_low-dimensional_2004,sykes_drag_2009}. For a weakly-interacting  Bose gas ($\gamma\ll 1$) this is not a severe restriction since the system size can be huge. Note that $1/L$ plays the role of an infrared cutoff for the momenta in the theory and that the final result for the induced interaction (see below) does not depend on it. Therefore, the obtained interaction is also valid in the thermodynamic limit.

Substituting the expansion (\ref{eq:wf_expw}) into Eq.~(\ref{eq:eomw}) we obtain the hierarchy of equations controlled by different powers of $\alpha$, which should be supplemented by the boundary condition (\ref{eq:BC}). The lowest order one,
\begin{align}
\left[-\frac{d^2}{2d X^2} +|\psi_0(X)|^2-1+\wt G\delta(X-R)\right]\psi_0(X)=0,
\label{eq:eom0w}
\end{align}
is known as the Gross-Pitaevskii equation \cite{gross_structure_1961,pitaevskii1961}. In the absence of the impurity (i.e., at $\wt G=0$), the latter mean-field equation has a simple real solution  $\psi_0(X)=\tanh X$ that satisfies the boundary condition $\psi_0(0)=0$ and has a vanishing gradient at infinity. At $\wt G\neq 0$ the mean-field boson density $\psi_0^2(X)$ becomes locally perturbed in the vicinity of the impurity. The solution of Eq.~(\ref{eq:eom0w}) in that case acquires the form
\begin{gather}
 \psi_0(X)=
 \begin{cases}
 \sqrt{\frac{1-2a}{1-a}}\,\text{sn}\left({\dfrac{X}{\sqrt{1-a}}};1-2a\right),&0\le X< R,\\
  \tanh(X+b),  &X> R.
 \end{cases}
 \label{eq:psisw}
\end{gather}
Here $\text{sn}$ denotes the Jacobi elliptic function, while $a$ and $b$ are the parameters to be determined from the conditions of the continuity (i) $\psi_0(R-0)=\psi_0(R+0)$ and the jump in the derivative (ii) $\psi_0'(R+0)-\psi_0'(R-0)=2\wt G\psi_0(R)$. Rather than solving the equations (i) and (ii) in full generality, let us consider the case $\wt G\ll 1$ in the following, where we can achieve a significant analytical progress. The latter inequality is equivalent to $G\ll g/\sqrt{\gamma}$, which also allows for $G$ that is much larger than $g$. Notice that at $a=b=0$, Eq.~(\ref{eq:psisw}) becomes the solution of Eq.~(\ref{eq:eom0w}) at $\wt G=0$. We then solve the equations (i) and (ii) at the leading order in small $\wt G$. One needs to expand the Jacobi sn function to the second order in small $a$ to obtain the correction to $\tanh X$. This yields
$a^2=4\wt G \sinh R/\cosh^3 R$, $b=-{\wt G}f(R)$,
where
\begin{align}
f(z)=\dfrac{\sinh R}{16\cosh^3R}(12z+8\sinh 2z+\sinh 4z).
\label{eq:f}
\end{align}
The latter expression gives
\begin{align}
\psi_0(X)=\tanh X-\dfrac{\wt G}{\cosh^2 X}\begin{cases} f(X),& 0\leq X \leq R,\\
	f(R),& X>R.
\end{cases}
\label{eq:psi0w}
\end{align}
Equation (\ref{eq:psi0w}) is valid for both, positive and negative  $\wt G$. It shows that the width of the Bose gas depletion due to the impurity is controlled by the healing length $\xi$, which at weak interaction is $\xi=\xi_\mu=1/n\sqrt\gamma$.

Let us now consider the effects of quantum fluctuations in the field operator (\ref{eq:wf_expw}), which is represented by the field operator $\hat\psi_1$. Its equation of motion is obtained from Eq.~(\ref{eq:eomw}) at order $\alpha$ and is given by
\begin{align}
i\partial_T \hat\psi_1=&\left(-\dfrac{\partial^2_X}{2}-1+2\psi_0^2\right)\hat\psi_1+\psi_0^2\hat\psi_1^\da  +\wt G\delta(X-R)\hat\psi_1.
\label{eq:gpedpsiw}
\end{align}
We seek the solution of Eq.~(\ref{eq:gpedpsiw}) in the form \cite{stringaribook} 
\begin{align}
\hat\psi_1(X,T)=\sum_k N_k \left[u_k(X)\hat b_k e^{-i\epsilon_k T}-v_k^*(X)\hat b_k^\da e^{i \epsilon_k T}\right].
\label{eq:uvw}
\end{align}
Here $N_k$ is a normalization factor, while the bosonic operators $\hat b_k$ and $\hat b_k^\da$ obey the standard commutation relation $[\hat b_k,\hat b_{q}^\da]=\delta_{k,q}$.

We first solve Eq.~(\ref{eq:gpedpsiw}) in the absence of the $\delta$ potential to linear order in $\wt G$, which enters through $\psi_0$ of Eq.~(\ref{eq:psi0w}). Substituting the ansatz (\ref{eq:uvw}) into Eq.~(\ref{eq:gpedpsiw}) yields the Bogoliubov-de Gennes equations for $u_k(X)$ and $v_k(X)$, which in terms of $S(k,X)=u_k(X)+v_k(X)$ and $D(k,X)=u_k(X)-v_k(X)$ become
\begin{align}
\epsilon_k S(k,X)={}&\left[-\dfrac{\partial^2_X}{2}+3\psi_0^2(X)-1\right]D(k,X), \label{eq:bdgSD2w}\\
\epsilon_k D(k,X)={}&\left[-\dfrac{\partial^2_X}{2}+\psi_0^2(X)-1\right]S(k,X). \label{eq:bdgSD1w}
\end{align}
With the help of Eq.~(\ref{eq:bdgSD1w}), Eq.~(\ref{eq:bdgSD2w}) becomes a fourth-order differential equation for $S(k,X)$. At $\wt G=0$, it has four independent solutions \cite{kovrizhin_exact_2001} $S_n(k,X)=(i k_n-2 \tanh X)e^{i k_n X}$, $n\in\{1,2,3,4\}$, while the energy dispersion is $\epsilon_k=\sqrt{k^2+k^4/4}$. The four roots of the energy dispersion that enter $S_n(k,X)$ are $k_{1,2}=\pm k$, $k_{3,4}=\pm i\sqrt{4+k^2}$ in terms of $k=\sqrt{2}\sqrt{\sqrt{\epsilon_k^2+1}-1}$. The solutions for $D_n(k,X)$ are obtained directly from $S_n(k,X)$ using  Eq.~(\ref{eq:bdgSD1w}).

At finite $\wt G$, we solve the fourth-order differential equation using the Bargmann method \cite{lamb}, where one is seeking the solution in the form $S_n(k,X)=P(k_n,X)e^{ik_n X}$, where $P(k_n,X)$ is the polynomial in $k_n$ of degree 5 in our case, with $X$-dependent coefficients. After dividing the obtained expression by $(2+k_n^2)(4+k_n^2)$ we eventually find
\begin{align}
S_n(k,X)={}&\biggl[ik_n-2\tanh X+\frac{\wt G\sinh R}{2\cosh^3 R}\biggl(\frac{3X}{\cosh^2 X}\notag\\
&-\frac{4k_n^2 X}{2+k_n^2}-\frac{4ik_n\cosh 2X}{4+k_n^2}+\frac{4-k_n^2}{4+k_n^2}\sinh 2X\notag\\
&+\frac{14+3k_n^2-8i k_n X}{2+k_n^2}\tanh X\biggr)\biggr] e^{i k_n X}
\label{eq:S<}
\end{align}
for $0\le X<R$ and
\begin{align}
S_n(k,X)=\left[ik_n-2\tanh X+\frac{2\wt G f(R)}{\cosh^2 X}\right]e^{i k_n X}
\label{eq:S>}
\end{align}
for $X>R$, where $f(R)$ is given by Eq.~(\ref{eq:f}).
The expressions (\ref{eq:S<}) and (\ref{eq:S>}) provide the solution of  Eq.~(\ref{eq:gpedpsiw}) without the $\delta$ potential and at linear order in $\wt G$ using the ansatz (\ref{eq:uvw}), Eq.~(\ref{eq:bdgSD1w}) and the relations $u=(S+D)/2$, $v=(S-D)/2$.

We can now account for the  $\delta$ potential  in the scattering problem (\ref{eq:gpedpsiw}). The solution for $S$ can be obtained as a linear combination of four independent solutions (\ref{eq:S<}) and (\ref{eq:S>}):
\begin{align}
S(k,X)=
\begin{cases}
\sum_{n=1}^4 t_n S_n(k,X),&0\leq X< R,\\
\sum_{n=1}^3 r_nS_n(k,X),&X>R,
\label{eq:Ssetupw}
\end{cases} 
\end{align}
with $r_2=1$. Equation (\ref{eq:Ssetupw}) describes an incoming wave from large $X$ that is partially reflected from the impurity at $X=R$ and that is fully reflected at the boundary $X=0$. In Eq.~(\ref{eq:Ssetupw}) we omitted from the linear combination the unphysical exponentially growing $S_4(k,X)$ in the region $X>R$.  Notice that there is a similar to Eq.~(\ref{eq:Ssetupw}) set of equations for $D(k,x)$ where $S_n$ is replaced by $D_n$, as follows from Eq.~(\ref{eq:bdgSD1w}).

The six coefficients that enter Eq.~(\ref{eq:Ssetupw}) are determined by the boundary conditions for the wave function. The condition (\ref{eq:BC}) implies $S(k,0)=D(k,0)=0$. This yields $t_2=t_1$ and $t_4=t_3$. The continuity of the wave function at the impurity position requires $S(k,R-0)=S(k,R+0)$, while  the jump in the derivative translates into $S'(k,R+0)-S'(k,R-0)=2\wt G S(k,R)$. Here the derivative is with respect to the second argument. There are two analogous equations for $D(k,X)$ function. Four conditions suffice to find the  remaining four coefficients: 
\begin{gather}
r_1=1+4 i\wt G\frac{k[k \cos (k R)-2 \sin (k R)]^2}{(2+k^2)(4+k^2)}, \label{eq:r1w}\\
t_1=1+2i\wt G\dfrac{k(k+2i) [k \cos (k R)-2 \sin (k R)]}{(2+k^2)(4+k^2)}e^{i k R},\\
r_3=4\wt G \dfrac{(2-\sqrt{4+k^2}) \left[k\cos (k R)-{2\sin (kR)}\right] }{k(2+k^2)(4+k^2)}e^{\sqrt{4+k^2} R},\\
t_3=-r_3 \frac{2+\sqrt{4+k^2}}{2-\sqrt{4+k^2}} e^{-2\sqrt{4+k^2} R}.
\end{gather}
They are expressed here in a simplified form where the limit $R\gg 1$ has been taken, while we also calculated them at any $R$ \footnote{See Supplemental Material for more details.}. However, in order to find the induced interaction on the impurity at separations longer than the healing length, it is sufficient to consider the large distance limit in the reflection and transmission amplitudes entering Eq.~(\ref{eq:Ssetupw}). The normalization of the solutions is obtained by requiring \cite{pitaevskii_bose-einstein_2003,walczak_exact_2011}  $N_k N_q\int dX (u_k u_q^*-v_k v_q^*)=\delta_{k,q}$, leading to $N_k=(\xi_\mu/4 L \epsilon_k)^{1/2}$. One can then verify that the bosonic commutation relation between the field operators are satisfied.

\section{Ground-state energy}

We are now in position to find the ground-state energy of the system. Since we work at constant chemical potential, we first consider the grand canonical energy
$E_\text{GC}=\la H\ra-\mu\int dx \la \hat\Psi^\da\hat \Psi\ra$ in the two leading orders at weak interaction. It can be expressed using the field decomposition (\ref{eq:wf_expw}) as
\begin{align}
E_\textrm{\tiny GC}=-\frac{\mu}{2\alpha^2}\int dX \psi_0^4-\frac{\mu}{2} \int d X \left[\left\langle ( i\partial_T \hat\psi_1^\dagger)\hat\psi_1 \right\rangle+\textrm{h.c.}\right].
\label{eq:Egc}
\end{align}
The second term in the right hand side can be further simplified into $-\mu \int dX \sum_k N_k^2|v_k|^2 \epsilon_k$ using the normal mode expansion (\ref{eq:uvw}). Equation (\ref{eq:Egc}) is derived with the help of the equations of motion (\ref{eq:eom0w}) and (\ref{eq:gpedpsiw}). We notice that $\hat\psi_2$ term of Eq.~(\ref{eq:wf_expw}) does not participate in the subleading term of Eq.~(\ref{eq:Egc}) as one can show using Eq.~(\ref{eq:eom0w}). 

The evaluation of the expression (\ref{eq:Egc}) is tedious. After performing the Legendre transformation to eliminate the chemical potential in favor of the density, we obtain the ground-state energy
\begin{align}
E={}&\frac{\hbar^2 n^3 L}{2m}\left(\gamma-\frac{4\gamma^{3/2}}{3\pi}\right) + \frac{\hbar^2 n^2}{2m}\left(\frac{4\sqrt\gamma}{3}-\frac{\gamma}{4}\right)\notag\\
&+G n \left[1-\frac{1}{\cosh^2 (r/\xi)}-\sqrt\gamma\,\, \mathcal{U}(r/\xi)\right].
\label{eq:E}
\end{align}
Here $\mathcal{U}(R)=1/16\pi R^2$ at large $R$ \cite{Note1}, 
while $\xi=1/n\sqrt\gamma$ is the healing length at weak interaction. The first term in Eq.~(\ref{eq:E}) is extensive and denotes the ground-state energy of the Bose gas with contact repulsion at two leading orders for $\gamma\ll 1$, which is in agreement with Bethe ansatz calculations \cite{lieb_exact_1963,popov_theory_1977}. The second term in Eq.~(\ref{eq:E}) is the boundary energy of the system, which represents the excess of energy due to the condition of vanishing density at the origin [cf.~Eq.~(\ref{eq:BC})]. At the leading order in weak interaction, this expression agrees with the perturbative Bethe ansatz result found in Ref.~\cite{gaudin_boundary_1971}. Here we have found the first correction $\propto\gamma$. 

The position-dependent part of the ground-state energy denotes the interaction potential energy between the impurity and the interacting Bose gas,
\begin{align}
U(r)=-G n \left[\frac{1}{\cosh^2 (r/\xi)}+\sqrt\gamma\,\, \mathcal{U}(r/\xi)\right].
\label{eq:U}
\end{align}
Equation (\ref{eq:U}) is our main result. As a consequence of broken translational invariance, $U(r)$ should be understood as the interaction between the impurity at position $r$ and the wall, representing the system boundary at the origin. The first term of Eq.~(\ref{eq:U}) denotes the classical mean-field interaction that follows from the solution (\ref{eq:psi0w}) of the Gross-Pitaevskii equation. It decays exponentially beyond the healing length $\xi$. One would then naively expect that the quantum correction $ \mathcal{U}(r/\xi)$ controlled by the small parameter $\sqrt\gamma$ can be neglected since it only introduces a small correction to the classical exponential result. The actual calculation reveals that $\mathcal{U}(r/\xi)$ indeed contains additional exponential corrections that we neglected. However, $ \mathcal{U}(r/\xi)$ also contains an important term that decays as a power law, leading to
\begin{align}
U(r)=-\frac{G n}{16K}  \left(\frac{\xi}{r}\right)^2
\label{eq:Ulr}
\end{align} 
at long distances $r\gg\xi$. Here we introduced $K=\pi/\sqrt{\gamma}\gg 1$, which denotes the Luttinger liquid parameter at weak interaction. The expression (\ref{eq:Ulr}) is the long-range interaction between the impurity and the wall that scales with the inverse square of the distance. Remarkably, the quantum-fluctuation correction term that is controlled by the small parameter $1/K$, becomes the dominant one at large distances, since it decays algebraically and thus overwhelms the exponential mean-field  contribution. The crossover scale where the two terms in Eq.~(\ref{eq:U}) are equal, 
\begin{align}
r_c\approx \xi\ln (8\sqrt{K}),
\end{align}
is of the order of $\xi$ and only weakly, i.e., logarithmically, depends on the interaction strength. At distances shorter than $r_c$ the interaction is exponential, $U(r)=-Gn/\cosh^2(r/\xi)$, while at large distances it crosses into a power law decay (\ref{eq:Ulr}). For a positive coupling constant $G$, the impurity is attracted toward the wall and vice versa. 

\section{Discussions}

How can our result (\ref{eq:Ulr}) be reconciled with the reported  $1/r^3$ decay \cite{schecter_phonon-mediated_2014} of the induced interaction between two impurities in Bose liquids? Let us consider the setup that consists of an impurity of infinite strength in the middle of the bosonic system at $x=0$. Such impurity creates an impenetrable potential for  quasiparticles. Thus the fluctuations in the two parts of the system, at $x<0$ and $x>0$ become uncorrelated, which results with $1/r^2$ interaction. This should be contrasted with $1/r^3$ law in the penetrable case when the impurity at $x=0$ is characterized by a finite strength. Viewed differently, an impurity characterized by any finite strength is an irrelevant perturbation in the renormalization group sense in quantum liquids that have the Luttinger liquid parameter $K>1$ \cite{kane_transport_1992}. However, our impurity of infinite strength cannot become irrelevant under the renormalization, since the effective tunneling term across such impurity is forbidden due to the impenetrability. Therefore, the  impurity (or equivalently the wall in another setup) causes slower decay of the induced interaction, which is revealed by our calculation.
 
Equation (\ref{eq:Ulr}) does not apply for an infinitely large $G$. In this case one obtains the Bose gas in a segment of finite length $r$ for which the asymptotic form of the induced interaction at large $r$ has a universal form 
\begin{align}\label{eq:Ustrong}
U(r)=-\frac{\pi\hbar v}{24r},
\end{align} 
where $v=\pi \hbar n/mK$ is the sound velocity. This result is quite general and applies to a massless scalar one-dimensional field, including bosons and fermions, with two infinitely strong $\delta$-function potentials   \cite{luscher_anomalies_1980,milton_casimir_2004,recati_casimir_2005}.

The interaction (\ref{eq:Ulr}) is obtained at zero temperature. However, it also applies at low temperature $T$, as long as the distance from the wall is smaller than the thermal length $\hbar v/2\pi T$. Although in this work we studied a static impurity, we point out that our main result (\ref{eq:Ulr}) will also hold for a slow mobile impurity. The corrections to Eq.~(\ref{eq:Ulr}) due to the impurity dynamics  will occur at higher order in the coupling strength $G$.

The scattering approach \cite{jaekel_casimir_1991} enables us to quantitatively understand the different forms of the Casimir-like interaction. In the latter theory, the interaction is expressed in terms of the reflection amplitudes $\mathbbm{r}_1(k)$ and $\mathbbm{r}_2(k)$ of the two impurities when they are considered individually, at the origin of the system. The induced interaction at large distances is given by
\begin{align}\label{eq:Uscatteringfull}
U(r)=\frac{\hbar v}{2\pi}\mathrm{Im}\int_0^\infty dk \ln\left[1-\mathbbm{r}_1(k) \mathbbm{r}_2(k) e^{2i k r}\right].
\end{align}
The latter expression can be further simplified in the case of small total reflection, $|\mathbbm{r}_1(k)\mathbbm{r}_2(k)|\ll 1$. The reflection amplitude of a weakly coupled impurity to the Bose gas [cf.~Eq.~(\ref{eq:Hw})] is $\mathbbm{r}_1(k)=-i Gk/2mv^2$ at $k\ll1/\xi$ \cite{reichert_casimir-like_2019}. In the case of two such impurities we have $\mathbbm{r}_1(k)=\mathbbm{r}_2(k)$ and therefore the expression (\ref{eq:Uscatteringfull})  gives $U(r)=-G^2 m\xi^3/32\pi \hbar^2 r^3$, which exactly matches the $1/r^3$ interaction previously obtained in Refs.~\cite{schecter_phonon-mediated_2014,reichert_casimir-like_2019,reichert_field-theoretical_2019}. In the case of a semi-infinite medium we must use $\mathbbm{r}_2=1$ and  Eq.~(\ref{eq:Uscatteringfull}) leads to our formula (\ref{eq:Ulr}). If, however, the impurity at the origin is coupled to the Bose gas by a finite coupling $G_2\gg \hbar v$, there is a crossover from $1/r^2$ to $1/r^3$ law that occurs at the distance $\sim \xi G_2/\hbar v$ \cite{reichert-unpublished}. Finally, in the case of infinitely coupled impurities to the Bose gas, i.e., at $\mathbbm{r}_1=\mathbbm{r}_2=1$,  Eq.~(\ref{eq:Uscatteringfull}) leads to Eq.~(\ref{eq:Ustrong}). We can conclude that the different scaling of the induced interaction is a consequence of the behavior of the reflection amplitudes of the isolated impurities.

The density-fluctuation induced interaction (\ref{eq:Ulr}) should be compared with the electrostatic one. For the impurity that is a neutral atom placed in a neutral background gas we thus need to estimate the van der Waals interaction. For two atoms, it scales with the sixth power of their inverse distance. For large separations of our impurity from the wall, $r\gg\xi$, we can estimate the electrostatic interaction on it by performing a pairwise summation with the background atoms that are in the region $[2r,\infty]$, since the contribution from the atoms that are in the region $(0,2r)$ approximately cancels. This leads to the nonretarded van  der Waals interaction $U_{\textrm{vdW}}(r)\propto 1/r^5$, that scales to zero much faster than our interaction (\ref{eq:Ulr}). In our other setup with a long system and an impurity at the origin that is very strongly coupled to it, the van  der Waals contribution would be zero due to symmetry reasons if we had no depletion of the Bose gas density around the origin. Therefore, the latter contribution arises from the local density depletion, which occurs  in the region of the characteristic width $\xi$. It leads to $U_{\textrm{vdW}}(r)\propto 1/r^6$, which is again negligible with respect to the interaction (\ref{eq:Ulr}).

For $G>0$, the attractive potential from the wall can produce bound states of the impurity. We estimate the kinetic energy of the impurity at separations of the order $\xi$ from the wall to be $\hbar^2/M\xi^2$, while its potential energy is $Gn$,  where we neglect the numerical factors of the order one. By $M$ is denoted the impurity mass. The potential energy is greater from the kinetic one at
\begin{align}\label{eq:boundstate}
G>\frac{m}{M}g,
\end{align} 
when the first bound state appears. A careful calculation \cite{reichert-unpublished} shows that the simple estimate (\ref{eq:boundstate}) is actually good. The quantum correction (\ref{eq:Ulr}) in the potential  produces   energy shifts of the bound state levels that are small due to the smallness of the coupling constant $G/K$. The bound state denotes the impurity localization near the end of the system. The latter phenomenon is reminiscent of self-localization of impurities in extended Bose-Einstein condensates \cite{cucchietti_strong-coupling_2006,sacha_self-localized_2006}. 

\section{Conclusions}

In conclusion, we have studied the induced interaction acting on the impurity in a semi-infinite one-dimensional interacting Bose gas. We found the induced long-range interaction (\ref{eq:Ulr}) of quantum origin. It scales quadratically with the inverse distance  from the wall and thus dominates the classical mean-field  exponential interaction at distances above the healing length. Our result (\ref{eq:Ulr})  scales much slower than the long-range van der Waals interaction and thus represents the dominant interaction on the impurity at long distances. A similar interaction mechanism should also exist in higher-dimensional systems. We also discussed the condition for the localization of the impurity near the wall. We finally notice that the dynamics of a particle moving in a $1/r^2$ potential is a potentially interesting problem to study, as it is predicted to exhibit fractal structure in the time domain \cite{gao_dynamical_2019}. 

\textit{Note added.} The first correction $\propto \gamma$ to the boundary energy in Eq.~(\ref{eq:E}) is in agreement with the exact result recently found in Ref.~\cite{reichert_exact_2019}.

This study has been partially supported through the EUR Grant No.~NanoX ANR-17-EURE-0009 in the framework of the ``Programme des Investissements d’Avenir".


%


\onecolumngrid
\newpage
\setcounter{equation}{0}
\setcounter{figure}{0}
\setcounter{section}{0}

\renewcommand{\theequation}{S\arabic{equation}}
\renewcommand{\thepage}{S\arabic{page}}
\renewcommand{\thesection}{S\arabic{section}}
\renewcommand{\thetable}{S\arabic{table}}
\renewcommand{\thefigure}{S\arabic{figure}}
\renewcommand{\bibnumfmt}[1]{[{\normalfont S#1}]}
\renewcommand{\citenumfont}[1]{S#1}

\begin{center}
	{\large\textbf{\mbox{Fluctuation-induced potential for an impurity in a semi-infinite one-dimensional Bose gas}}
		\\\vskip 5pt
		\normalsize{Supplemental material}
	}
	
	Benjamin Reichert, Aleksandra Petkovi\'{c}, and Zoran Ristivojevic\\
	\vskip 1mm
	
	\textit{Laboratoire de Physique Th\'{e}orique, Universit\'{e} de Toulouse, CNRS, UPS, 31062 Toulouse, France}
\end{center}
\vskip 0mm

\section{Reflection and transmission amplitudes}

Here we give the expressions that we obtained for the reflection and transmission amplitudes at arbitrary $R$:
\begin{align}
t_1={}&t_2=1+ \frac{i \wt G}{(2+k^2)(4+k^2)}\biggl\{\left[2(2+k^2)^2-8(1+k^2)e^{2i kR} \right]\dfrac{\tanh^2R}{k}\notag\\
& +2i\left[(5+2k^2)e^{2i kR}-\dfrac{14+7k^2+k^4}{2+k^2}-i kR(4+k^2) \right]\tanh R\notag\\
& -2i\left[5e^{2i kR}-\dfrac{14+7k^2+k^4}{2+k^2}-i kR(4+k^2) \right]\tanh^3 R  +\dfrac{2+4k^2+k^4}{k}(e^{2i kR} -1)+\dfrac{6\tanh^4R}{k}(e^{2i kR}-1)\biggr\},\\
t_3={}&t_4=-\dfrac{4\wt Ge^{-\sqrt{4+k^2}R}}{k(2+k^2)(4+k^2)}\biggl\{ \sin(k R) +\left[k \cos (kR)+\sqrt{4+k^2}\sin (kR)\right]\dfrac{2\tanh R}{2+k^2} \notag\\
&+\left[k(1+k^2)\cos (kR)-(3+k^2)\sqrt{4+k^2}\sin (kR) \right]\dfrac{2\tanh^3R}{2+k^2}\notag\\
& -3\sin (kR) \tanh^4R  +\left[k\sqrt{4+k^2}\cos (kR)-2\sin (kR)\right] \tanh^2R    \biggr\},\\
r_1={}&1- \dfrac{4i\wt G}{(2+k^2)(4+k^2))}
\biggl\{\left[\left(5+2 k^2\right) \sin (2 k R)-k \left(4+k^2\right) R\right]\tanh R   -\left[\left(2+k^2\right)^2-4 \left(1+k^2\right) \cos (2 k R)\right]\dfrac{\tanh ^2R}{k}\notag\\ &+\dfrac{6\sin^2(kR)}{k}\tanh^4R+(2+4k^2+k^4)\frac{ \sin ^2(k R)}{k}+\left[k R(4+k^2)-5\sin (2kR) \right]\tanh^3R\biggr\},\\
r_3={}&\dfrac{8\wt G}{k(2+k^2)(4+k^2)}\biggl\{  \sin( k R)      \sinh(\sqrt{4+k^2}R)-3\sin (kR) \sinh (\sqrt{4+k^2}R) \tanh^4R \notag\\
&+\left[k \cos(kR)\sinh(\sqrt{4+k^2}R)-\sqrt{4+k^2} \sin (k R)\cosh (\sqrt{4+k^2}R) \right]\dfrac{2\tanh R}{2+k^2}\notag\\
&-\left[ k\sqrt{4+k^2} \cos (kR)\cosh (\sqrt{4+k^2}R)+2\sin (kR)\sinh( \sqrt{4+k^2}R)\right]\tanh^2R \notag\\
&+\left[(3+k^2)\sqrt{4+k^2}\sin (kR)\cosh(\sqrt{4+k^2}R)+k(1+k^2)\cos (kR) \sinh (\sqrt{4+k^2}R) \right]\dfrac{2\tanh^3 R}{2+k^2}\biggr\}.
\end{align}
To obtain the leading term of amplitudes presented in the main text at $R\gg 1$ simply amounts to replace $\tanh R\to  1$. In the main text, some other simplifications have also been done. In order to calculate the subleading term of the grand canonical energy at order $\wt G$ which is $\propto\wt G\int dX\int dk  \text{Re}\{v^0_k v^1_k\}$, where we use the notation $ v_k=v^0_k+\wt G v^1_k+O(\wt G^2)$, we dismiss the terms which after integration over $X$ would produce the exponential contributions in $\exp (-\sqrt{4+k^2}R)$. For $X>R$, for example, the product $r_3 S_3$ contains terms $\propto \exp [-\sqrt{4+k^2}(R+X)]$ which after integration over $X$ produces a term $\propto\exp (-2\sqrt{4+k^2}R)$. Therefore the term $\propto \exp (-\sqrt{4+k^2}R)$ can be safely set to 0 in $r_3$. 
The same is done for $0\leq X\leq R$ where $t_3$ is $\propto\exp (-\sqrt{4+k^2}R)$ and multiplied by $S_3$ which produces a term $\propto \exp (-\sqrt{4+k^2}X)$.  After integration over $X$, the latter produces terms $\propto\exp (-\sqrt{4+k^2}R)$ and $\propto \exp (-2\sqrt{4+k^2}R$) and therefore we can set $t_3=0$ (but we must keep the nonzero form for $t_4$, despite the exact relation $t_4=t_3$).

\section{Some details about $\mathcal{U}(R)$}

The function $\mathcal{U}(R)$ defined by Eq.~(21) of the main text, at large $R$ has the form
\begin{align}\label{eq:UU}
\mathcal{U}(R)={}&\frac{1}{\pi}\int_0^\infty dx \left[\frac{4-5x^2}{(4+x^2)R}-2x^2\right]\frac{\sin(2Rx)}{(4+x^2)^{3/2}}\notag\\ ={}&-\frac{2}{\pi}+(4R-1)\left[I_0(4R)-\textbf{L}_0(4R)\right] -\left(4R-\frac{1}{2}\right)\left[I_1(4R)-\textbf{L}_{-1}(4R)\right],
\end{align}
where we neglected exponentially decaying contributions. In Eq.~(\ref{eq:UU}), 
$I_\nu(R)$ denotes the modified Bessel function of the first kind, while  $\textbf{L}_{\nu}(R)$ is the modified Struve function. We notice the expansion
\begin{align}
\mathcal{U}(R)=\frac{1}{16\pi R^2}\left[1+\frac{1}{R}+\frac{15}{16R^2}+O\left(\frac{1}{R^3}\right)\right].
\end{align}

\end{document}